\documentstyle[psfig]{mn2e}

\title[Early type galaxies colours and star formation histories]
{Optical/near-infrared colours of early-type galaxies and 
constraints on their star formation histories}
\author[P. A. James et al.]
{P. A. James$^{1}$, M. Salaris$^{1}$, J. I. Davies$^{2}$, 
S. Phillipps$^{3}$ and S. Cassisi$^{4}$\\ 
$^{1}$Astrophysics Research Institute, Liverpool John Moores
University, Twelve Quays House, Egerton Wharf, Birkenhead CH41 1LD, UK\\
$^{2}$School of Physics and Astronomy, University of Cardiff,
The Parade, Cardiff CF24 3YB, Wales, UK\\
$^{3}$Department of Physics, University of Bristol, Royal Fort,
Tyndall Avenue, Bristol BS8 1TL, England, UK\\
$^{4}$INAF-Osservatorio Astronomico di Collurania, Via Mentore Maggini,
I-64100 Teramo, Italy}
\begin{document}

\date{}

\pagerange{\pageref{firstpage}--\pageref{lastpage}} \pubyear{2005}

\maketitle

\label{firstpage}

\begin{abstract}

We introduce and discuss the properties of a theoretical
$(B-K)$-$(J-K)$ integrated colour diagram for single-age,
single-metallicity stellar populations. We show how this combination
of integrated colours is able to largely disentangle the well known
age-metallicity degeneracy when the age of the population is greater
than $\sim$ 300~Myr, and thus provides valuable estimates of both age
and metallicity of unresolved stellar systems. We discuss in detail
the effect on this colour-colour diagram of $\alpha$-enhanced metal
abundance ratios (typical of the oldest populations in the Galaxy),
the presence of blue horizontal branch stars unaccounted for in the
theoretical calibration, and of statistical colour fluctuations in low
mass stellar systems. In the case of populations with multiple stellar
generations, the luminosity-weighted mean age obtained from this
diagram is shown to be heavily biased towards the youngest stellar
components.  We then apply this method to several datasets for which
optical and near-IR photometry are available in the literature.  We
find that LMC and M31 clusters have colours which are consistent with
the predictions of the models, but these do not provide a sensitive
test due to the fluctuations which are predicted by our modelling of
the Poisson statistics in such low-mass systems. For the two Local
Group dwarf galaxies NGC185 and NGC6822, the mean ages derived from
the integrated colours are consistent with the star formation
histories inferred independently from photometric observations of
their resolved stellar populations.  

The methods developed here are applied to samples of nearby early-type
galaxies with high quality aperture photometry in the literature.
A sample of bright field and Virgo cluster elliptical galaxies is
found to exhibit a range of luminosity-weighted mean ages  from 3 to
14~Gyr, with a mean of $\sim$8~Gyr, independent of environment, 
and mean metallicities at or just above the solar value.
Colour gradients are found in all of the galaxies studied, in the
sense that central regions are redder.  Apart from two radio galaxies,
where the extreme central colours are clearly driven by the AGN, and
one galaxy which also shows a radial age gradient, these
colour changes appear consistent with metallicity changes at a
constant mean age.  Finally, aperture data for five Virgo early-type dwarf
galaxies shows that these galaxies appear to be shifted to lower mean
metallicities and lower mean ages (range 1 to 6 Gyr) than their higher
luminosity counterparts.
\end{abstract}


\section{Introduction}

Within the currently favoured $\Lambda$CDM models of galaxy and large
scale structure formation, dwarf galaxies play a crucial role. The
first stars are predicted to form in small dark matter halos of mass
$\approx 10^{6}$ $M_{\odot}$ at redshifts of $z \approx 20-30$ (see
Bromm and Larson 2004). These small dark matter halos subsequently
form the building blocks from which the larger galaxies we see around
us today are assembled (Kauffman et al. 1993; Cole et al. 2000;
Mathis et al. 2002). As with every building site many of the smaller
component parts are left behind once the larger structures are formed
and so the $\Lambda$CDM model predicts that there should be many small
dark matter halos around us today.

If this debris of small dark matter halos can be associated with dwarf
galaxies then there is a large discrepancy between what is observed
and what is predicted by theory, this is known as the sub-structure
problem (Moore et al. 1999; Klypin et al. 1999). The standard
$\Lambda$CDM model, without dwarf galaxy formation suppression
processes, predicts many more small dark matter halos (with a low mass
slope of the mass function of $\alpha \approx -2$) compared to
observations of the faint end of the global luminosity function of
galaxies ($\alpha \approx -1.2$; Blanton et al. 2001, Norberg et al.,
2002). There have been numerous suggestions for a solution to this
problem. These typically involve the prevention of gas falling into
small dark matter halos (Efstathiou 1992), the removal of gas via
supernova winds from the first generation of stars (Dekel and Silk
1986) or the `squelching' of dwarf galaxy formation by ionising
photons (Tully et al. 2002). These mechanisms prevent or severely
inhibit star formation in small halos; the halos still exist, but they
remain hidden from us. Interest in the possiblity that star formation
in dwarf galaxies has in some way been delayed compared to larger
galaxies has come from the extensive studies made using Sloan Digital
Sky Survey data. This seems to indicate that it is the low luminosity
galaxies that contain the youngest stars (Kauffmann et al., 2003). It
is difficult to see how this `downsizing' of the galaxy population
fits in with the standard $\Lambda$CDM model (Kodama et al. 2004), but
it is an observation we can test with the methods we discuss in this
paper. Recently other solutions have been suggested, for example by
adjusting the initial dark matter fluctuation spectrum, something
suggested by combining the WMAP, 2dF and Lyman-$\alpha$ forest
observations (Spergel et al. 2003), small dark matter halos themselves
become rare.

By studying in detail the sub-structure problem in different galaxy
environments we can hope to gain an insight into the mechanisms that
prevent dark matter halos revealing themselves to us as dwarf galaxies
or comment on how universally rare small halos might be. We have
previously quantified the relative numbers of dwarf galaxies in
different environments and find many more dwarf galaxies in the Virgo
and Fornax clusters than can be found in the general field (Phillipps 
et al. 1998a; Sabatini et al. 2003, 2004; 
Roberts et al. 2004, 2005; Davies et al. 1988; Kambas et al. 2000). 
The general field result is also consistent with
observations of the dwarf galaxy population of the Local Group (Mateo
1998; Pritchet \& van den Bergh 1999). It appears that global star
formation suppression mechanisms do
not work. The environment plays a major role in either producing a
large number of additional dwarf galaxies or in revealing more of the
existing dark matter halos to us.

The expected properties of the first galaxies to form within the
$\Lambda$CDM model seem to be more closely matched to the properties
of globular clusters (a star formation event followed by almost total
gas expulsion) than it does to the more complicated star formation
histories of dwarf galaxies. Although dwarf galaxies
appear to be relatively simple stellar systems (there is little
evidence for spiral stellar density waves, bars or jets, but see
Barazza et al. 2002) they can be more susceptible to exterior influences.
In the main dwarf galaxies show a morphology density relation with the
gas poor dSph galaxies being closer to the bright galaxies in the
Local Group and closer to the centres of galaxy  
clusters than are the dIrr galaxies (Ferguson \& Sandage 1989;  
Mateo 1998; Boyce et al. 2001; Sabatini et al. 2003;
Roberts et al. 2004). Within a cluster environment, or when they are
close to a more massive galaxy, dwarf galaxies may be subject to ram
pressure and/or tidal stripping, tidal destruction or `harassment'
processes (for a review see Sabatini et al. 2004).

Currently we essentially only have detailed star formation histories
for Local Group dwarf galaxies because these are the only ones for
which we have resolved observations of their stellar populations.
These have been interpreted using stellar colour magnitude diagrams
and models that implement different stellar ages and metallicities
(Grebel \& Stetson 1999; Smecker-Hane et al. 1996; Gallart et al.
1999; Mighell and Rich 1996; Skillman et al. 2003). 
Each of the Local Group dwarf galaxies
has a quite different star formation history when studied in detail,
with stellar populations covering a wide range of ages and metallicities. In
general the dSph galaxies have no young stars; the stellar population
is dominated by stars that are older than 10 Gyr, but there are some
intermediate age stars (1-10 Gyr) indicating that star formation
continued for many Gyr after their formation (Grebel 2002). The
current, most widely used, $\Lambda$CDM solution to the substructure
problem (that small galaxies in the early Universe should experience a
single catastrophic episode of star formation) is not supported by
observations of the star formation histories of Local Group dwarf
galaxies or by the large numbers of dwarf galaxies found in clusters.

Although subject to similar physical processes dwarf galaxies in
clusters may have very different origins and histories to those in the
Local Group. Within the cluster environment there are many more dwarf
galaxies per giant (Phillipps et al. 1998b, Sabatini et al. 2003),
so the clusters are not being assembled from groups like the Local
Group. Also unlike the Local Group there are many early type (dE)
dwarf galaxies that are not companions of the giant galaxies (Sabatini
et al. 2003, 2004; Roberts et al. 2004). Ideally we would like to
study the star formation histories of these galaxies in the same
detail as has been done for the Local Group galaxies to see if they
are consistent with each other and with the properties predicted by
$\Lambda$CDM - predominantly old, gas poor, low metallicity
galaxies. Currently this is not possible for dwarf galaxies in the two
nearest clusters to us (Virgo and Fornax) for two reasons. Firstly,
because the bulk of the stellar population cannot be resolved at these
distances (although the resolution of galaxies at these distances is a
key science driver for the proposed Extremely Large
Telescopes). Secondly, many of the dwarf galaxies we are particularly
interested in are of such low surface brightness (LSB) that it is impossible to obtain spectra
and derive line indices (Worthey 1994, Vazdekis and Arimoto 1999,
Kuntschner 2000). Thus, at present we are forced to use population
indicators that can be applied to completely unresolved LSB stellar
populations.

In the present study we explore the use of optical/near-IR colours as
diagnostics of luminosity-weighted mean ages and metallicities of
stellar populations.  We demonstrate that such
colours have similar diagnostic power to spectral line indices, while
being much less demanding in terms of observing time, data reduction and 
analysis. The development of
such indicators is particularly timely with the recent release of the
full Two-Micron All-Sky Survey (Jarrett et al. 2000; henceforth
2MASS), and the prospect of deeper surveys such as the United Kingdom
Infrared Deep Sky Survey (UKIDSS) recently commenced with the
United Kingdom Infrared Telescope (UKIRT).

Model predictions of optical/near-IR colours have previously been
presented by several groups (e.g. Fioc and Rocca-Volmerange 1997;
Smail et al. 2001, Girardi et al.  2002; Bruzual and Charlot 2003) and
observational studies have already highlighted the potential of such
measurements for breaking the age-metallicity degeneracy, which is the
first requirement of any successful population diagnostic method.
Peletier, Valentijn \& Jameson (1990) used $(U-V)$, $(B-V)$ and $(V-K)$ colours to
constrain the ages and metallicities of 12 elliptical galaxies, while
Peletier \& Balcells (1996) demonstrated that $(U-R) - (R-K)$
colour-colour plots gave useful information on the mean ages of galaxy
bulges and disks.  Bell et al. (2000) and Bell \& de Jong (2000) used
$(B-R)$ and $(R-K)$ colours in the analysis of the star formation
histories of low surface brightness and spiral galaxies respectively.
Puzia et al. (2002) and Hempel \& Kissler-Patig (2004) have recently
demonstrated the power of $(V-I) - (V-K)$ colour-colour diagrams
applied to analysis of the star-formation histories of unresolved
extragalactic globular clusters.

In this paper we first develop our own predictions of the
optical/near-IR colours of stellar populations as a function of age
and metallicity, using models developed by two of the authors (MS and
SC).  We identify the $(B-K)$-$(J-K)$ colour-colour plane as a
particularly sensitive one for breaking age-metallicity degeneracy,
and use photometry to test these predictions for systems (globular
clusters and two dwarf galaxies in the Local Group) with existing
independent measurements of age and metallicity.  We also apply these
methods to a small number of Virgo cluster elliptical and dwarf
elliptical galaxies which have sufficiently high quality photometry in
the 2MASS catalogues and $B$-band photometry in matched apertures, in
the first stage in a study of the star formation history of cluster
galaxies.

\section{The theoretical calibration}

We employed the stellar evolution model library by Pietrinferni et
al.~(2004), that spans the metallicity (scaled solar metal
distribution) range between Z=0.0001 and Z=0.04 (10 different
metallicities in total); this model grid enables one to compute
isochrones for ages from $\sim$ 20-40~Myr up to 15-20~Gyr, covering
all evolutionary phases from the Zero Age Main Sequence (ZAMS) up to
the beginning of the Thermal Pulse (TP) phase along the Asymptotic
Giant Branch (AGB) or carbon ignition, depending on the mass of the
most evolved objects. In order to compute correctly the near-IR
integrated colours of stellar populations older than $\sim$100~Myr we
have extended the evolutionary tracks of the masses that evolve along
the AGB phase, until the end of the TP phase. This has been performed
by means of the synthetic AGB evolution technique (see e.g. Marigo et
al.~1996 and references therein) starting from the stellar structures
at the beginning of the TPs. The use of this technique suffices for
our purposes and avoids extremely time-consuming evolutionary
computations. The analytical relationships necessary for the synthetic
AGB computations (e.g. an equation for the rate of growth with time of
the carbon-oxygen core, the core mass-luminosity relationship, etc)
are taken from Wagenhuber \& Groenewegen~(1998); they are obtained
from full evolutionary AGB calculations covering the relevant range of
initial masses and metallicities. The analytical formulae for the
determination of the model $T_{eff}$ along the extended AGB phase are
not given explicitly in Wagenhuber \& Groenewegen (1998), but can be
found in Wagenhuber~(1996); the mass loss rate along the TP phase is
computed according to the methods of Vassiliadis \& Wood~(1993). The
synthetic AGB evolution is stopped when the models start to cross the
Hertzsprung-Russell diagram towards their white dwarf cooling
sequence.

Using this extended set of evolutionary tracks we computed isochrones
for ages running from 40~Myr to 14~Gyr, suitably transformed to
various photometric filters using the bolometric corrections and
colour transformations described in Pietrinferni et al.~(2004). For
the AGB TP phase we have used the transformations described in Girardi
et al.~(2002) modified in the carbon star regime to reproduce the
empirical $(B-K)-(J-K)$ colour-colour relationships for Galactic and LMC
clusters' carbon stars, as obtained from data by Mendoza \&
Johnson~(1965) and Frogel \& Cohen~(1982).

Integrated magnitudes and colours for the simple stellar populations
(SSPs) described by our isochrones have been determined using a
Salpeter~(1955) Initial Mass Function (IMF). We adopted as lower and
upper IMF mass cutoffs the values 0.1$M_{\odot}$ and 100$M_{\odot}$,
respectively, as in Bruzual \& Charlot~(2003). The choice of these
cutoffs is important for the determination of the total mass of the
system and the integrated magnitudes, but it is not critical for the
integrated colour-colour calibration discussed in this paper, due to
the following two reasons. First, the integrated colours are
independent of the total stellar mass, apart from the case of stellar
populations with masses typical of star clusters (see next
subsection). Only in this case does the value of the total mass of the
stellar population play a role in the determination of the integrated
colours.  Second, the upper limit of the mass contributing to the
total flux and colours is fixed by the most massive star evolving at
the age of the population, and it is at most around 10$M_{\odot}$ in
our discussion. As for the lower mass limit, we have found that in the
photometric filters of interest, MS objects less massive than
$\sim$0.5$M_{\odot}$ do not provide an appreciable contribution to the
integrated flux. To give a quantitative estimate of the contribution
of these low-mass stars, we have recomputed the $(B-K)$ and $(J-K)$
colours neglecting stars with masses below 0.5$M_{\odot}$. We found
that the $(J-K)$ values are decreased by less than 0.01 mag and
$(B-K)$ is decreased by only 0.03-0.05 mag over the full range of ages and
metallicities of our calibration.

\begin{figure}
\psfig{file=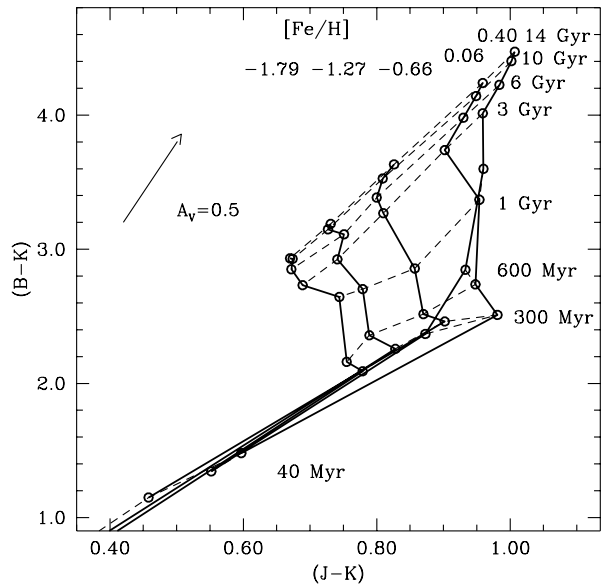,width=8.3cm}
\caption{Theoretical integrated $(B-K)$ and $(J-K)$ colours for
isochrones with ages of 40~Myr, 300, and 600~Myr,1, 3, 6, 10 and
14~Gyr (open circles), and metallicities [Fe/H] from $-$1.79 (leftmost
solid line) to +0.40 (rightmost solid line). Dashed lines connect
points corresponding to SSPs of constant age.  The arrow indicates the
direction of the reddening vector.}
\label{fig1}
\end{figure}

Figure~\ref{fig1} displays the theoretical integrated $(B-K)$ and
$(J-K)$ colours for a subset of our isochrones with ages of,
respectively, 40, 300 and 600~Myr, and 1, 3, 6, 10, and 14~Gyr (from bottom to
top) and (from left to right) [Fe/H]=--1.79, --1.27, --0.66, 0.06,
and 0.40. This combination of colours is able -- for ages larger than
$\sim 600$~Myr -- to appreciably disentangle the effect of age and
metallicity on the observed integrated fluxes. Some residual
degeneracy is still present between 600~Myr and 1~Gyr at the high
metallicity end of the calibration.  The integrated colour $(J-K)$
appears to be a good tracer of the initial metallicity of a Simple
Stellar Population (SSP) and is weakly affected by age, whereas the
integrated $(B-K)$ is a good age indicator, mildly affected by the SSP
metallicity. The direction of the reddening vector is also shown in
the figure.

We have studied the contribution of the individual evolutionary phases
to the $B$, $J$ and $K$ integrated magnitudes and found, not
unexpectedly, that the $J$ and $K$ integrated fluxes are dominated by
AGB stars when the SSP age is below $\sim 1$~Gyr, and by upper Red
Giant Branch (RGB) objects for higher ages. As for the integrated $B$
flux, the main contribution comes always from the upper Main Sequence
(MS) and TO stars. This means that the integrated $(J-K)$ colour is
mainly determined by the colour of AGB and/or RGB stars, whose
location is strongly affected by their initial metallicity, whereas
$(B-K)$ is sensitive to the magnitude and colour of the TO, hence to
the SSP age.

It is also very interesting to notice how the theoretical $(B-K)-(J-K)$
integrated colour diagram closely mirrors the behaviour of the
theoretical calibrations of pairs of Lick indices (e.g. Worthey~1994)
like, for example $H_{\beta}-Fe5270$, where $H_{\beta}$ is a tracer of
the stellar population age and $Fe5270$ is sensitive mainly to the
metal content.

\begin{figure}
\psfig{file=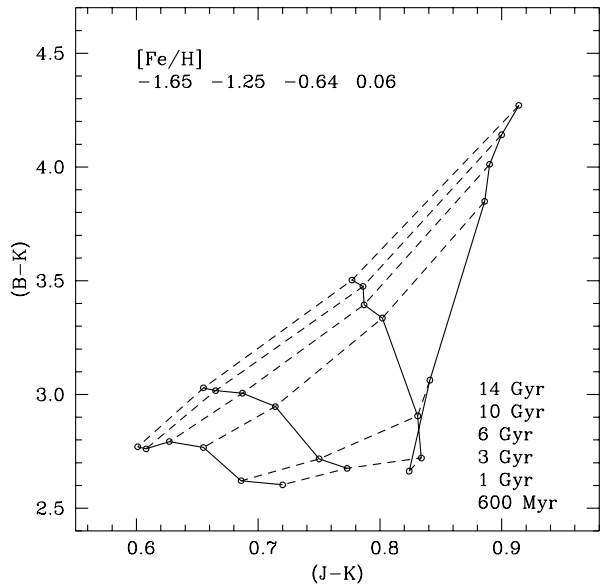,width=8.3cm}
\caption{Theoretical integrated $(B-K)$ and $(J-K)$ colours from
Girardi et al.~(2002) models. The calibration is shown for the age
range where the age-metallicity degeneracy is broken.}
\label{fig2}
\end{figure}

For ages above $\sim$10~Gyr the $(B-K)$ colour tends to lose
sensitivity to age, whereas below $\sim$300~Myr there is full
degeneracy between age and metallicity, since the lines corresponding
to different metallicities overlap.  These general properties of the
integrated $(B-K)$-$(J-K)$ diagram are not peculiar to our set of
models; they are displayed also i.e., by the Girardi et al.~(2002) and
Bruzual \& Charlot~(2003) integrated colours. As an example, we show
in Fig.~\ref{fig2} the Girardi et al.~(2002) theoretical diagram, for
the age range where the age-metallicity degeneracy is broken.

\subsection{The effect of statistical fluctuations}

Given that the fast evolving upper RGB and AGB stars are the main
contributors to the $(J-K)$ integrated colours, it is very important to
assess the effect of small number statistics on the integrated $(B-K)$
and $(J-K)$ colours. When the mass of the SSP under scrutiny is small --
total masses up to $10^5-10^6~M_{\odot}$,
the upper end of the mass spectrum of star clusters -- the number of
stars in these fast evolutionary phases is subject to sizable
fluctuations from one SSP to another, with potentially large
fluctuations of the integrated magnitudes and colours. To quantify
this effect we have performed a series of Monte-Carlo simulations. 

We have considered the following combinations of ages ($t$) and [Fe/H]
values: [Fe/H]=$-$1.79 and $t=$10~Gyr, [Fe/H]=$-$0.66 and $t=$10~Gyr,
[Fe/H]=$-$1.27 and $t=$600~Myr.
For each one of these combinations, we have drawn stars 
randomly from a Salpeter IMF and
placed them in their evolutionary phases along the isochrone at that
age, until a cluster mass of $10^5~M_{\odot}$ is reached. The 
integrated colours of the resulting populations have been then computed.
In general, we found that the fluctuations are negligible in $B$ since this
wavelength range is dominated by the much more populous MS phase.

\begin{figure}
\psfig{file=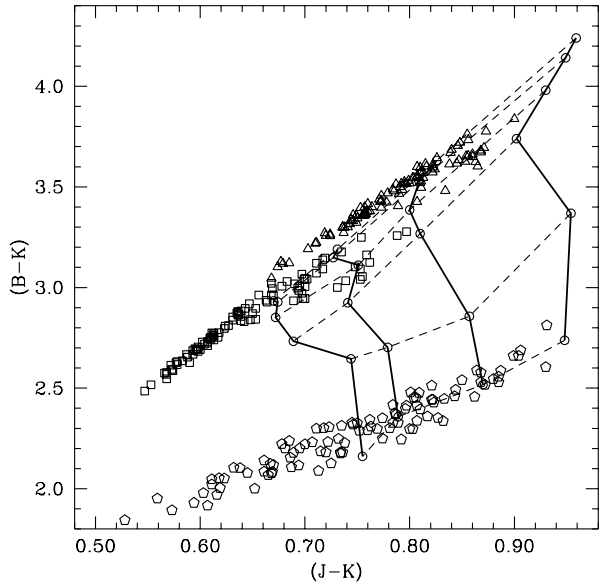,width=8.3cm}
\caption{Model colours with overlaid points showing the results of 100
simulations of 10$^5 M_{\odot}$ synthetic stellar
populations. The scatter in the colours of the synthetic populations
stems from the statistical
fluctuations in the numbers of the most luminous stars. 
Open squares, triangles and pentagons display populations with, respectively: 
[Fe/H]=$-$1.79 and t=10~Gyr, [Fe/H]=$-$0.66 and t=10~Gyr,
[Fe/H]=$-$1.27 and t=600~Myr.}
\label{fig3}
\end{figure}

The open squares, triangles and pentagons in Fig.~\ref{fig3} 
display the integrated colours obtained
from 100 realizations each for the three adopted age and metallicity
combinations. The $(J-K)$ colours
show a very large spread due to statistical fluctuations of the number
of stars along the upper RGB and AGB phases. This spread increases for
decreasing age because of the shorter timescales. Hence, there are larger
number fluctuations along the AGB phase that dominate the $(J-K)$
colours at 600~Myr. The colour spread causes a 3$\sigma$ uncertainty
of $\approx$2~dex in the inferred metallicity at 600~Myr, and 
$\approx 1$dex at 10~Gyr. The fluctuation of $(B-K)$ is entirely due to
the fluctuation of the $K$ magnitudes and it is interesting to notice
that the path in the colour-colour plane described by the 100
realizations follows a vector that is almost parallel to lines of
constant age. At 600~Myr the fluctuations cause an almost negligible
uncertainty on the age, while the effect is more important at 10~Gyr,
due to the lower sensitivity of $(B-K)$ to age in this age range.

\begin{figure}
\psfig{file=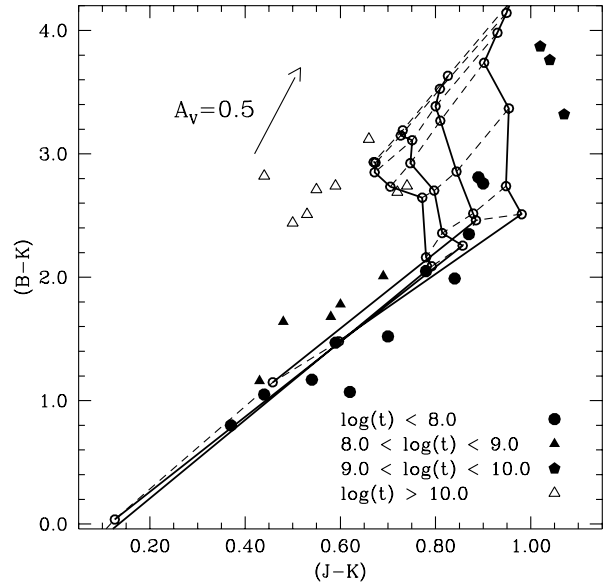,width=8.3cm,height=8.3cm}
\caption{Observed colours of LMC clusters (points) overlaid on our
model predictions for [Fe/H] between $-$1.79 and +0.06. Different
point types denotes the mean cluster ages
as determined by van den Bergh~(1981).  
The effect of an additional
extinction $A_V$=0.5 mag is also displayed.}
\label{fig4}
\end{figure}

\begin{figure}
\psfig{file=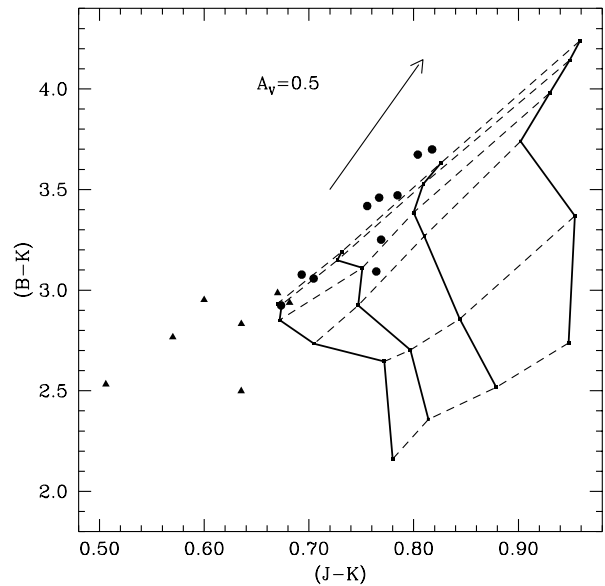,width=8.3cm,height=8.3cm}
\caption{Observed colours of M31 globular clusters overlaid on our
model predictions for [Fe/H] between $-$1.79 and +0.06. Filled triangles denote clusters with [Fe/H]$\leq
-$1.60, while filled circles display clusters with higher
metallicities.}
\label{fig5}
\end{figure}

These fluctuations can clearly be seen in Figs.~\ref{fig4} and
\ref{fig5}, which show the colours of bright star clusters in the
Large Magellanic Cloud (LMC) and M31 globular clusters, superimposed
on our models.  The $B$ photometry for the LMC clusters is taken from
van den Bergh~(1981), the $J$ and $K$ photometry (plus reddening
estimates) is from Persson et al.~(1983); the age classification comes
from van den Bergh (1981), and is based on integrated $UBV$
colours. Figure~\ref{fig4} shows that the young LMC clusters scatter
along the tracks predicted for the early colour evolution of clusters,
and illustrates that at these ages ($<$300~Myr) the colours predicted
by the models are completely degenerate.  The oldest clusters lie
fairly close to the colour predicted for an age of 14~Gyr and [Fe/H] =
$-$1.79, but with a large scatter very similar in size and direction
to that shown in the simulated old clusters in Fig.~\ref{fig3}.

In case of the M31 globular clusters we considered the sample
recently observed by Rich et al.~(2005). The bright resolved stars of
this cluster sample provide an estimate of the individual photometric 
metallicities and reddenings; the integrated colours are taken from
Galleti et al.~(2004). The metallicity range spanned by these M31 clusters
is approximately enclosed by the two metallicities [Fe/H]=$-$1.79 and
$-$0.66 used in our simulations of 10~Gyr old populations.
Figure~\ref{fig5} shows clearly how the
observed colour distribution follows to a good approximation the distribution of
points in Fig.~\ref{fig3} for the two old populations. 

As an aside, we have also computed Monte-Carlo simulations for
the $(V-K)$ colour fluctuations, employing 
the same cluster masses and metallicities as 
an analogous test performed by Bruzual \& Charlot~(2003), and 
recovered essentially the same results as these authors.

To conclude, Figs.~\ref{fig4} and \ref{fig5} generally
show strong encouragement for the predictions of our modelling; but
it is also clear that this statistical fluctuation effect seriously limits the
use that can be made of stellar clusters in testing these predictions.
It should be emphasised that these statistical
fluctuations of the integrated colours are only significant for stellar
systems up to the sizes of large globular clusters, and are completely
negligible for galaxies of 10$^8 M_{\odot}$ and larger (i.e. all
galaxies considered in the current paper).

\subsection{Anomalous Horizontal Branch colours}

In general, colours and spectral features involving the blue part of
the spectrum can be seriously affected by the presence of blue
Horizontal Branch (HB) stars not accounted for in the theoretical
calibration. Such stars have high temperatures and can add an
appreciable contribution to the flux e.g. in the $B$ filter and Balmer
lines. The models we used in our analysis have been computed with a
standard Reimers~(1975) mass loss law along the RGB, with the free
parameter $\eta=0.2$. This choice allows one to reproduce the mean
colour of the HB in Galactic globular clusters of various
metallicities, that qualitatively corresponds to a shift from red to
blue -- for a mean value of the age of $\sim$ 12-13 Gyr -- when the
metallicity decreases. However, as has been well known for many years (e.g.
Sandage \& Wildey 1967) some Galactic globular clusters of
intermediate metallicity (e.g. [Fe/H] between $\sim -$1.6 and $\sim  -$1.1) like the pairs
M3-M13, NGC362-NGC288, show significantly different mean colours (M13 and
NGC 288 having a much bluer HB with respect to their counterparts,
that display an HB colour in line with other clusters of the same
metallicity) in spite of having apparently the same age and the same
[Fe/H]. This occurrence is the so-called second parameter phenomenon.
Indeed, one explanation for this different morphology would be a
higher age for the blue HB clusters (lower evolving mass hence bluer
HB colours) but this appears not to be confirmed by their TO
brightness.

\begin{figure}
\psfig{file=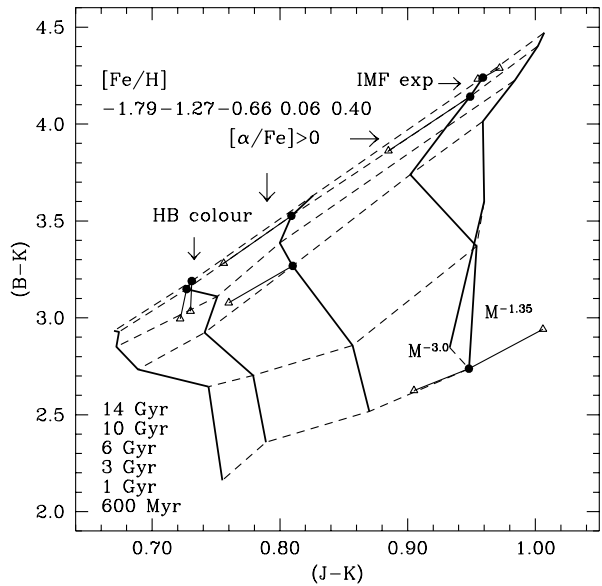,width=8.3cm}
\caption{Model colours, with added points (triangles) showing the
effect of, respectively, a bluer Horizontal
Branch (for a population with t=10~Gyr and 14~Gyr, [Fe/H]=$-$1.27) an
$\alpha$-enhanced metal mixture 
(for intermediate-old populations with [M/H]=$-$0.66 and 0.06), and 
a variation of the IMF exponent
(for two representative populations with [Fe/H]=0.06 and ages of
600~Myr and 14~Gyr, respectively). 
Filled circles mark the integrated colours
predicted by our standard calibration for the age-metallicity
combinations used in our experiments. The solid lines connect
the standard colours with those obtained after accounting for a 
bluer Horizontal Branch,  an $\alpha$ element enhancement, or a change of
the IMF exponent.}
\label{fig6}
\end{figure}

The effect of an unaccounted-for blue HB on the $(B-K)$-$(J-K)$ integrated
colours has been evaluated with the following numerical test. For ages 
of 10 and 14~Gyr, and [Fe/H]=$-$1.27 we have computed isochrones using a
more efficient mass loss along the RGB (the value of $\eta$ has been
doubled) that produced an HB much bluer than the case of our standard
colours discussed above. The integrated colours for these blue HB
populations have then been computed homogeneously with our standard
case, and are displayed in Fig.~\ref{fig6}. As expected, the effect on the
inferred age is significant; these blue HB populations appear to be $\sim
6$~Gyr old when the age is evaluated using our standard calibration,
instead of 10 and 14~Gyr. The metallicity estimate is only slightly affected,
because of a small contribution of the HB to the J and K colours.
This quantitative estimate is typical for all populations with ages
greater than $\sim$10~Gyr, i.e. the age range where 
this effect is found at work in Galactic clusters.

\subsection{The effect of an $\alpha$-enhanced metal distribution}

The scaled solar metal distribution, used in almost all stellar
evolution computations and adopted in our standard models, is not
universal. Probably the most relevant example of this phenomenon is
the fact that the initial chemical composition of both the old field
and globular cluster stars in the Galactic halo displays a ratio
[$\alpha$/Fe]$\sim0.3-0.5$ (Carney 1996), where with $\alpha$ we
denote the so-called $\alpha$ elements (O, Ne, Mg, Si, S, Ca, Ti).
Also elliptical galaxies (e.g. Worthey, Faber \& Gonzalez~1992) and
metal poor stars in other Local Group galaxies appear to have been
formed with an initial [$\alpha$/Fe]$>$0, although with different
degrees of enhancement.

We have studied the effect of an $\alpha$-enhanced metal mixture on
the integrated $(B-K)$ and $(J-K)$ colours by performing the following
test. The [$\alpha$/Fe]=0.4 models by Cassisi et al.~(2004) extended
along the AGB TP phase have been employed to determine the integrated
colours for the following age-metallicity combinations: [M/H]=$-0.66$,
age 3 and 10~Gyr; [M/H]=0.06, age 10~Gyr. We recall that in the case
of a scaled solar metal mixture [M/H]=[Fe/H], whereas in case of this
adopted $\alpha$-enhanced mixture [Fe/H]=[M/H]$-$0.35. The results of
this test are displayed in Fig.~\ref{fig6}.  The age inferred for the
$\alpha$-enhanced populations when using our standard calibration is
the correct one for the 10~Gyr population, whereas there is an age
overestimate of about 1-2~Gyr for the 3~Gyr case. The
[$\alpha$/Fe]$>$0 $(J-K)$ colours are shifted to the blue with respect
to the scaled solar ones, and their values are within $\sim$0.1~dex of
the $(J-K)$ corresponding to [Fe/H]=[M/H]$-$0.35 and a scaled solar metal
mixture.  We have found these same results concerning the recovered
age and [Fe/H] when we performed the same comparisons at both lower
and larger metallicities. Overall, the combined effect of an
$\alpha$-element enhancement on both the theoretical models (i.e.,
evolutionary timescales and the luminosity-$T_{eff}$ plane) and the
bolometric corrections to the $B$, $J$ and $K$ bands, makes this
colour combination a good tracer of the [Fe/H] ratio, even in the
presence of a ratio [$\alpha$/Fe]$>$0.

To summarize, a scaled solar calibration of the $(B-K)$-$(J-K)$ diagram is
able to recover the correct age and approximately the correct [Fe/H]
of an $\alpha$-enhanced population with [$\alpha$/Fe] typical of the
Galactic halo. In case of a younger 3~Gyr old population, the inferred
age tends to be overestimated by 1-2~Gyr.
Similar offsets are found over the entire range of metallicities
investigated in this paper. 

\subsection{The effect of the IMF exponent}

To estimate the effect of possible departures from the Salpeter
IMF, we have computed the integrated
colours of solar metallicity SSPs using different exponents of the IMF. 
The results for two specific ages (14~Gyr and 600~Myr,
respectively) are shown in Fig.~\ref{fig6}. 
As a general rule, a more negative value of the exponent (MS dominated
population) shifts the integrated colours towards bluer values, 
whereas the opposite is true
in case of a less negative exponent (giant dominated population). 

The reason for this behaviour is easily understood by recalling that
MS objects less massive than $\sim0.5M_{\odot}$ do not contribute
appreciably to the total integrated flux in the our photometric bands
of choice, even if we decrease the exponent of the IMF significantly.
For example, if we set the exponent to $-$3, excluding the
contribution of stars with $M < 0.5M_{\odot}$ decreases the $(J-K)$
colour of a 10~Gyr, [Fe/H]=0.06 population by only 0.012~mag and the
$(B-K)$ colour by 0.08~mag.  An MS dominated SSP will have a larger
contribution of MS stars to the integrated flux, but MS objects in the
relevant mass range ($M > 0.5M_{\odot}$) are typically bluer than
bright RGB and AGB objects (that still provide the largest
contribution to the $J$ and $K$ fluxes) and thus make the integrated
colours slightly bluer compared to the case of the Salpeter exponent.
The opposite is true in a giant dominated SSP.  The magnitude of this
effect increases with decreasing age, because the extension of the MS
reaches brighter magnitudes and bluer colours when the age decreases.

\subsection{Composite stellar populations}

The colour-colour diagram of Fig.~\ref{fig1} refers to simple stellar
populations, made of coeval stars born with the same chemical
composition. In the case of a complex star formation history, the age and
metallicity obtained from this diagram will reflect some mean value,
that is determined by the number and luminosity of the objects belonging
to various stellar generations formed during the system's lifetime.

The age sensitive $(B-K)$ colours are obviously very sensitive to the
presence of young populations, whose Main Sequence is very bright in
the B band. The metallicity-sensitive $(J-K)$ colour is strongly affected by
intermediate age populations (300 -- 600~Myr old) that show a very
bright AGB.

We found that, keeping the metallicity constant, the age inferred from
$(B-K)$ is always biased towards the youngest stellar component. In the
case, for example, of two star formation bursts 40~Myr and 10~Gyr
old, the $(B-K)$ is completely dominated by the youngest population,
even if this contributes to the total system mass by only $\sim$~3-5\%.  This
is very similar to the conclusions reached by Tantalo \& Chiosi (2004)
in a theoretical study of the age-determinations using H$\beta$ and [MgFe] line
indices. Their models showed that a burst of star formation involving
2\% of the stellar mass of an otherwise old population would
significantly perturb the derived age to young values for a period of
order 2~Gyr after the onset of the burst.

A powerful test for the adequacy and usefulness of the theoretical
$(B-K)-(J-K)$ calibration discussed in this paper, is provided by the
comparison of our model predictions with integrated colours of Local
Group galaxies, whose star formation histories (star formation rate -
SFR - plus age-metallicity relation - AMR) are independently estimated
from their resolved stellar populations.

2MASS $J$ and $K$ total integrated magnitudes for the Local Group
galaxies NGC~6822 (dwarf Irregular) and NGC~185 (dwarf Elliptical)
have been taken from Jarrett et al.~(2003), while the integrated $B$
magnitudes have been taken from the Third Reference Catalogue of
Bright Galaxies (de Vaucouleurs et al.~1991). Estimates of the star
formation histories of these two galaxies, using
Colour-Magnitude-Diagrams of their resolved stellar populations, are
available in the literature (i.e., Gallart et al.~1996, Wyder~2001 for
NGC~6822; Mart\'inez-Delgado, Aparicio \& Gallart~1999 for NGC~185).
Observed $(J-K)$ colours, and $K$ magnitudes have been transformed to
the Bessell \& Brett~(1988) system using the transformations by
Carpenter~(2001). Extinction corrections from Schlegel,
Finkbeiner \& Davis~(1998) have been applied, and no correction for
possible internal extinction has been considered.

\begin{figure}
\psfig{file=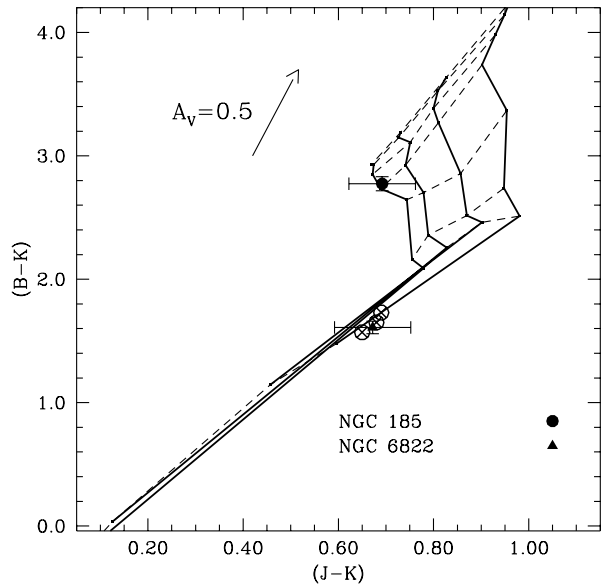,width=8.3cm,height=8.3cm}
\caption{Observed colours of the Local Group galaxies 
NGC~185 and NGC~6822 overlaid on our
model predictions. Crossed circles denote theoretical integrated
colours obtained from the SFR and AMR of NGC~6822, as estimated by
Wyder~(2001 -- see text for details).}
\label{fig7}
\end{figure}

Figure~\ref{fig7} displays the integrated colours of these two galaxies
compared to our model grid. NGC~6822 appears to be young, located in 
the region of full degeneracy between age and metallicity
effects. In light of our previous discussion about composite stellar
populations, the position of NGC~6822 on our colour-colour diagram 
is fully consistent with its estimated star formation
history, that displays a sizable star formation rate at young
ages. NGC~185 appears to be older, with a mean age
slightly above 3~Gyr and a mean metallicity of the order of
[Fe/H]$\sim -$1.8. This is again consistent with the star formation
rate inferred from its resolved stellar population, dominated by ages
above $\sim$ 8~Gyr, and a much smaller star formation activity at
young ages.

A more quantitative test has been performed in case of NGC~6822 (an
analogous quantitative estimate for NGC~185 could not be performed,
because of the lack of a firm estimate of the AMR for this galaxy).
Wyder~(2001) provides the SFR and AMR for three fields within the
central bar, and two outer fields. The star formation history of the
central bar fields -- located in a region that contributes most to the
total integrated magnitude of the galaxy -- has been employed to
determine their expected integrated colours. These theoretical
predictions, displayed in Fig.~\ref{fig7}, show a very good agreement
with the observed integrated colour of the whole galaxy.

\section{Application of this method to early-type galaxies}

In this section we will use the model colours to determine properties
of the early-type galaxy population in the nearby Virgo cluster and
the general field.  The 2MASS database contains $JHK$ photometry for
most of the giant galaxies in Virgo, and a smaller but useful number
of the brighter dwarf galaxies, so the main problem is finding
matching $B$-band photometry. Our initial analysis made use of $B_T$
extrapolated total magnitudes from the Virgo Cluster Catalogue
(Binggeli, Sandage \& Tammann 1985), which we combined with the 2MASS $K_{ext}$
magnitudes to derive total $B-K$ magnitudes (see e.g. Mannucci et
al. 2005).  However, we found systematic offsets between the colours
thus obtained and those calculated from aperture photometry of the
same galaxies (Poulain 1988, Poulain\& Nieto 1994), in the sense that
the extrapolated magnitudes gave bluer colours than aperture
measurements, typically by 0.2--0.3~mag., so we restricted the
analysis to galaxies with matched aperture photometry in all three
passbands ($B$, $J$ and $K$).

One extremely useful database is that of Michard (2005) who studies
the optical/near-infrared colours of 94 elliptical galaxies out to a
distance modulus $(m-M)_0$=33.52 and with total integrated $B$
magnitude $B_T <$18.8, using 2MASS near-IR images and optical aperture
photometry from the literature.  He quotes colours within an effective
aperture $A_e$; in every case this is interpolated from the published
multi-aperture photometry in exactly the same way for the optical and
near-infrared photometry, so no systematic colour differences should
result from this procedure.  Note that the $B$ photometry we use,
which is originally from Poulain(1988) and Poulain \& Nieto (1994), is
not mentioned in the main text of Michard's paper, but is available in
the associated on-line data.  The integrated magnitudes have been
dereddened and $k$-corrected as described in Michard~(2005), and the
infrared magnitudes and colours transformed to the standard Bessell \&
Brett~(1988) system following Carpenter~(2001).

\begin{figure}
\psfig{file=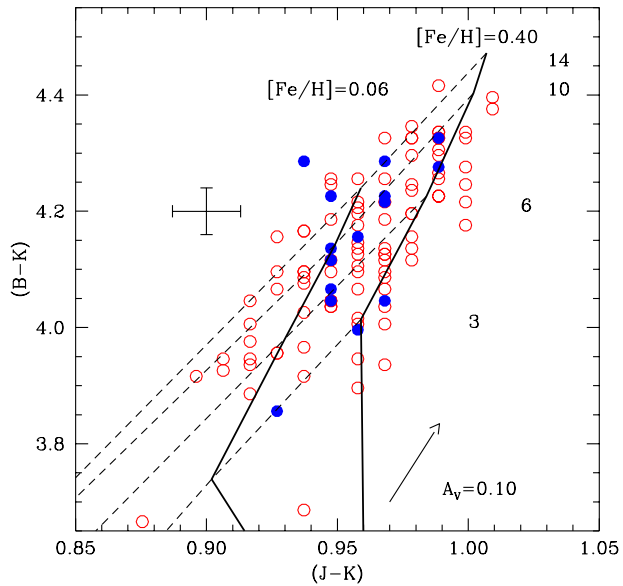,width=8.3cm,height=8.3cm}
\caption{Integrated colours of a sample of nearby elliptical galaxies overlaid 
onto the theoretical calibration.  Selected ages and metallicities are marked; 
the typical observational (1$\sigma$) errors and the reddening vector are also
shown.  Filled circles show the galaxies belonging to the Virgo cluster.
}
\label{fig:michard}
\end{figure}

Figure~\ref{fig:michard} displays the colours of the 94 galaxies overlaid
onto the theoretical calibration. The filled symbols show 16 objects
belonging to the Virgo cluster, the highest density environment in the
region of space sampled by these data. The distribution of about 99\% 
of the galaxies in the sample follows nicely a mean 
linear relationship between
$(B-K)$ and $(J-K)$, corresponding to an average age of 8~Gyr and a range
in [Fe/H] values between +0.4 and about the solar value.
Superimposed on this mean relationship is an age
spread that spans the range between $\sim$3 and $\sim$14 Gyr.
As an example, in the Virgo cluster NGC~4478 has a luminosity weighted
age of about 3~Gyr, whereas NGC~4374 is about 10~Gyr old.
These age and metallicity spreads appear to be real, and 
are not due to the observational errors, which are quite small in comparison.
We note that the use of the models of Girardi et al. (2002) would result in 
somewhat higher mean metallicities and younger mean ages in the colour range
covered by this galaxy sample. 

The distributions of points corresponding to the Virgo galaxies and
the rest of the objects are statistically the same, as verified with a
2-dimensional Kolmogorov-Smirnov (KS) test (Press et al.~1992). Thus
we find no evidence for environmental effects on the
luminosity-weighted mean ages and metallicities of bright ellipticals
in the sample of Michard (2005).

In general, the integrated $(B-K)-(J-K)$ colours of this sample show
that, if ellipticals formed in a single star formation episode, their
epoch of formation spans a large range of redshifts.  
If multiple generations are present together with an old
component formed soon after the Big Bang, the age range derived in
this analysis can be explained in terms of different fractions of
intermediate age/young populations, which bias the luminosity-weighted
ages towards lower values.  The presence of very young populations
(ages below 100~Myr) at a level larger than $\sim$1-2\% (by mass) can
be excluded by the results of this colour-colour analysis.

\section{Age and metallicity gradients within galaxies}

\begin{figure}
\psfig{file=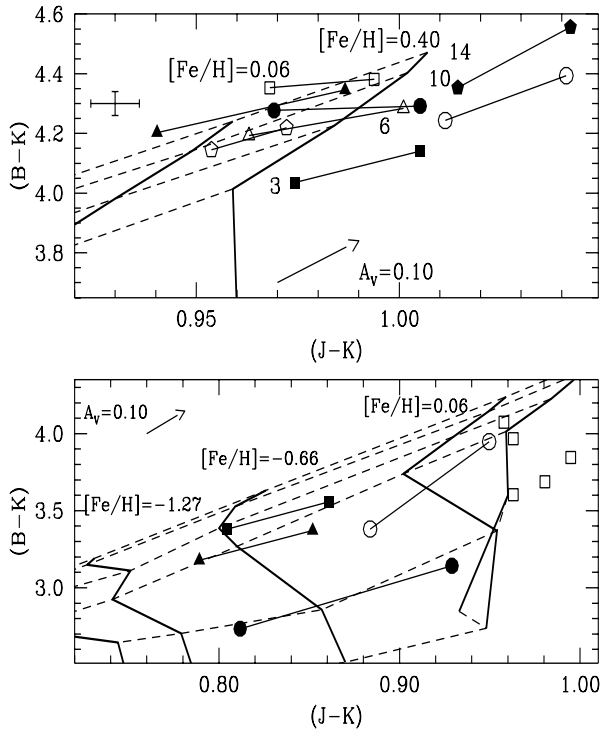,width=8.3cm,height=10.3cm}
\caption{Radial colour gradients for selected galaxies.  The upper plot 
shows 8 bright Virgo cluster elliptical galaxies, which are identified
in the text.  In each case, the
range of aperture colours is indicated by the two points plotted for
each galaxy, with the right-hand point indicating the colours in the
smallest aperture and the left hand point those for the largest aperture.
The lower plot shows similar results for the 5 dwarf galaxies, also
identified in the text.  Again, the two points indicate the range of
colours found, with the reddest colours being generally found for the
smallest apertures.  The clear exception to this rule is NGC~4640, for
which all available aperture colours are plotted explicitly.  In this
case, the smallest aperture (10$^{\prime\prime}$ diameter) has the
reddest $B-K$ colour, and this colour gets monotonically bluer to the
largest aperture (50$^{\prime\prime}$).  However, there is no overall
trend in $J-K$ for NGC~4640.}
\label{fig:galapcol}
\end{figure}

One complication in comparing results from integrated colours with
literature studies based on line indices is that the latter are
generally based on spectroscopic measurements which are only available
for central regions of the galaxies concerned.  Thus it is important
to check for systematic gradients in derived properties of stellar
populations which may affect such comparisons.  One advantage of the
methods presented here is that it is simple to get spatially-resolved
information from array imaging, and hence to compare central and
global stellar populations.

A subsample of 8 Virgo galaxies with aperture photometry from
Poulain~(1988), Poulain \& Nieto~(1994) and 2MASS is displayed in
Fig.~\ref{fig:galapcol}, upper frame; the sample contains NGC~4261
(open circles), NGC~4365 (filled circles), NGC~4374 (open pentagons),
NGC~4472 (open squares), NGC~4473 (filled squares), NGC~4486 (filled
pentagons), NGC~4621 (filled triangles) and NGC~4636 (open
triangles). The aperture sizes (diameters) used by these authors were 
$21\farcs9$, $31\farcs2$, $43\farcs4$, $60\farcs9$ and
$86\farcs6$ (although in the case of NGC~4473 the largest aperture
is $60\farcs9$).  However, for every one of these galaxies we found
a monotonic trend in colours with aperture, with the reddest colours
being found for the smallest aperture, so we only plot symbols for the
smallest and largest apertures. The typical error bars associated to
the colours through an individual aperture are displayed in the upper
left corner.

The general trend is of redder colours for decreasing aperture, a well
known result (see, e.g. Peletier et al.~1990).  The two galaxies with
the reddest $(J-K)$ colours are NGC~4261 and NGC~4486, both of which
are radio galaxies with strong nuclear emission, and so the colours
are not reliable indicators of the underlying stellar population.
Interestingly, for NGC~4374, NGC~4473, NGC~4636 and, to some extent,
also NGC 4621 (if one takes into account the size of the
error bars on the colours of each individual aperture) the colour
gradients correspond to a sequence of increasing metallicity at
approximately constant age. The slope of these radial colour gradients
in the $(B-K)-(J-K)$ plane is therefore similar to the mean slope of
the total integrated colours for the whole sample of galaxies, a
result in agreement with Peletier et al.~(1990) who employed the
$(U-V)-(V-K)$ plane. Approximately constant luminosity-weighted ages
at different radii are also found in relatively young looking galaxies
like NGC~4636 and NGC~4473 (luminosity weighted ages of $\sim$6~Gyr
and $\sim$3~Gyr respectively); this implies that any intermediate
age/young population that eventually coexists with an old
($\approx$10-14~Gyr) first stellar generation must be evenly spread
throughout the parent galaxy.

The situation for NGC~4365, and to some extent NGC~4472, is
different. Moving from the largest aperture towards the smallest one,
the $(B-K)$ colour stays approximately constant, whereas $(J-K)$
increases. This implies that the luminosity-weighted age decreases
from $\sim$ 10~Gyr to below 6~Gyr, and [Fe/H] increases from
$\sim0.06$ to above 0.40.  Internal extinction should not be
responsible for these different colour gradients (nor for the colour
gradients of all other objects displayed in the figure) given that
these are all early-type galaxies.  Only a progressively bluer
morphology (which would have to be bluer than the underlying
theoretical calibration) of the HB with decreasing radius would be
able to keep the NGC~4365 $(B-K)$ colour constant for increasing
$(J-K)$ and metallicity.

The lower frame in Fig. \ref{fig:galapcol} shows similar results for 5
dwarf galaxies in the Virgo cluster. These are the only Virgo cluster
galaxies classified as dwarfs by Binggeli et al. (1985) which have
2MASS photometry with errors in $J-K <$ 0.08~mag., and for which
$B$-band photometry is available from the survey of Sabatini et
al. (2003).  The latter study covered two strips of total area 25
deg$^2$ within the Virgo cluster.  The galaxies plotted are UGC~7436,
type dE5, plotted as filled squares; IC~781, type dS0,N, plotted as
filled triangles; IC~3292, type dS0, plotted as filled circles;
UGC~7399A, type dE5,N, plotted as open circles; and NGC~4640, type
dS0,N, plotted as open squares.  Again, only the smallest
(10$^{\prime\prime}$) and largest (40--60$^{\prime\prime}$) apertures
are plotted for most of these; where this is done, the smallest
apertures always have the reddest colours.  Only in the case of
NGC~4640 have all the aperture colours been plotted separately.  For
some reason this galaxy exhibits very different colour gradients from
the other dwarfs.  In this case the $B-K$ colours redden monotonically
from the smallest aperture (10$^{\prime\prime}$) to the largest
aperture (50$^{\prime\prime}$), while the $J-K$ colour does a peculiar
loop to redder colours and back again.  Apart from this galaxy, the
overall colour gradients appear to be dominated by increasing
metallicity towards the centres of these galaxies, with any age
gradients being small, and those which are seen going in both
directions.  However, the most interesting result from this small
number of dwarf galaxies is not in the colour gradients, but in the
overall position of these dwarfs compared to the more luminous
galaxies seen both in the upper frame of Fig. \ref{fig:galapcol} and
in Fig. \ref{fig:michard}.  The dwarfs are shifted with respect to the
brighter galaxies, both towards lower metallicities and, more
interestingly, towards lower luminosity-weighted mean ages, with a range
for the dwarfs of 1--6~Gyr.  It is hard to draw any general
conclusions from such a small sample, but the difference is quite
striking, particularly given that we are studying the very brightest
end of the dwarf galaxy luminosity function, because of the difficulty in
detecting lower luminosity and hence lower-surface-brightness galaxies
with the relatively shallow 2MASS data.  

Thus we do see significant gradients in colours as a function of
aperture size, which needs to be borne in mind when comparing
integrated properties using colours with nuclear properties using line
indices, particularly since spectroscopic apertures will tend to be
much smaller even than the smallest apertures used here.

\section{Discussion and conclusions}

We will summarise briefly our conclusions on the nature of the
stellar populations of the various galaxy samples as revealed by their
optical/near-infrared colours.  It is useful to remember at this
point that the ages and metallicities derived are luminosity-weighted
mean values, and we have shown that, as for spectroscopic indices,
the results can be heavily skewed by the presence of even a small
(a few per cent by mass) young stellar population in an otherwise old galaxy.

For the bright elliptical galaxy sample of Michard (2005), we find
that the integrated colours within the effective radius for these
galaxies reveal stellar populations with mean ages averaging about
8~Gyr, but with a real scatter, larger than the measurement errors,
covering the range from 3--14~Gyr.  Mean [Fe/H] values are just above
solar, but again there is a real scatter, from just below solar to
just above [Fe/H]$=$0.4.  The Virgo cluster galaxies contained within
Michard's sample show no statistically significant differences in
their integrated parameters than the sample as a whole, so there is no
evidence for strong environmental effects on either mean stellar ages
or metallicities.

A more detailed study of population gradients within 8 Virgo cluster
giant galaxies and 5 Virgo cluster dwarf galaxies reveals a strong
trend for central regions of all galaxies to have higher metallicities
than the galaxies as a whole.  Evidence for age gradients is much more
marginal, but one giant galaxy shows evidence for a younger central
population, while the dwarfs show weak evidence for age gradients in
both directions.

Considering the 5 Virgo dwarf galaxies together, we find that they
have younger mean stellar ages than their more luminous counterparts.
Clearly, this is too small a sample to be drawing general results
about the dwarf galaxy population as a whole, but it should be noted
that such a result would be consistent with recent observational
results by Caldwell et al. (2003), using spectral indices of
low-luminosity galaxies in Virgo and lower density environments, and
Kauffmann et al. (2003), who perform a statistical study of the star
formation histories of galaxies from the Sloan Digital Sky
Survey.  

There are two conclusions we can make from this preliminary
study. Firstly, if we assume that all these galaxies contain a simple
stellar population (SSP), then quite contrary to the hierarchical
model of galaxy formation the smallest galaxies have not formed first
at redshifts of 6-10 (ages greater than 12~Gyr for standard
$\Lambda$CDM) but, this does fit in with the `downsizing' scenario
discussed in the introduction. Secondly, if these galaxies do not
contain a SSP but do have a small fraction of young stars then the
feedback mechanisms discussed in the introduction are inefficient -
these galaxies have retained at least some gas to enable them to
continue to form stars over long periods of time (see also the model by
Kobayashi, 2005, which has this property).  This second
interpretation fits in very well with observations of the resolved
stellar populations found in Local Group dwarf galaxies (Grebel \&
Stetson 1999; Smecker-Hane et al. 1996; Gallart et al.  1999; Mighell
\& Rich 1996) and discussed in the introduction.

Unfortunately, it is only possible to study the brightest of the Virgo
dwarf elliptical galaxies using 2MASS photometry.  However, with new
data becoming available, e.g.  from the UKIDSS project, it should be
possible to derive the required colours for some hundreds of Virgo
cluster dwarfs, pushing significantly further down the luminosity
function.

\section*{Acknowledgments}

This publication makes extensive use of data products from the Two
Micron All Sky Survey, which is a joint project of the University of
Massachusetts and the Infrared Processing and Analysis
Center/California Institute of Technology, funded by the National
Aeronautics and Space Administration and the National Science
Foundation. We warmly thank Tom Jarrett for his very prompt
calculation of some additional aperture photometry which was not
available in the online 2MASS database, and Rory Smith for providing 
optical magnitudes for Virgo cluster dwarfs.  This research has made use of
the NASA/IPAC Extragalactic Database (NED) which is operated by the
Jet Propulsion Laboratory, California Institute of Technology, under
contract with the National Aeronautics and Space Administration.  The
referee is thanked for several helpful comments, one of which led to
the discovery of the problem with the use of extrapolated total
magnitudes in calculating $B-K$ colours.

\section*{References}



Barazza F., Binggeli B., Jerjen H., 2002, AA, 391, 823\\ 
Bell E. F., Barnaby D., Bower R. G., de Jong R. S., Harper D. A.,
Hereld M., Loewenstein R. F., Rauscher B. J., 2000, MNRAS, 312, 470\\
Bell E. F., de Jong R. S., 2000, MNRAS, 312, 497\\
Bessell M. S., Brett J. M., 1988, PASP, 100, 1134\\
Binggeli B., Sandage A., Tammann G., 1985, AJ, 90, 1681\\
Blanton et al., 2002, AJ, 121, 2358\\
Boyce P. J., Phillipps S., Jones J. B., Driver S. P., Smith R. M.,
Couch W. J., 2001, MNRAS, 328, 277\\
Bromm V., Larson R., 2004, ARAA, 42, 79\\
Bruzual G., Charlot S., 2003, MNRAS, 344, 1000\\
Caldwell N., Rose J. A., Concannon K. D., 2003, AJ, 125, 2891\\
Carney B.W., 1996, PASP, 108, 900\\
Carpenter J. M., 2001, AJ, 121, 2851\\ 
Cassisi S., Salaris M., Castelli F., Pietrinferni A., 2004, ApJ, 616,
498\\
Cole S., Lacey C., Baugh C., Frenk C., 2000, MNRAS, 319, 168\\
Davies J., Phillipps S., Cawson N., Disney M., Kibblewhite E., 1988, MNRAS, 232, 239\\
de Vaucouleurs G., de Vaucouleurs A., Corwin H. G., Buta R. J., Paturel
G., Fouqu\'e P.,  1991, ``Third Reference Catalogue of Bright Galaxies'',
Springer-Verlag\\
Dekel A., Silk J., 1986, ApJ, 303, 39\\
Efstathiou G., 1992, MNRAS, 256, 43\\
Ferguson H. C., Sandage A., 1989, ApJ, 346, 53\\
Fioc M., Rocca-Volmerange B., 1997, AA, 326, 590\\
Frogel J. A., Cohen J. G., 1982, 253, 580\\
Gallart C., Aparicio A., Bertelli G., Chiosi C., 1996, AJ, 112, 2596\\ 
Gallart C, Freedman W., Aparicio A., Bertelli G., Chiosi C., 1999,
AJ, 118, 2245\\
Galleti S., Federici L., Bellazzini M., Fusi Pecci F., Macrina S.,
2004, A\&A, 416, 917\\
Girardi L., Bertelli G., Bressan A., Chiosi C., Groenewegen M. A. T.,
Marigo P., Salasnich B., Weiss A., 2002, A\&A, 391, 195\\
Grebel E., 2002, in `Modes of Star Formation and the Origin of Field
Populations', ASP conf. ser.,
Vol. 285, eds. E. K. Grebel and W. Brandner\\
Grebel E., Stetson P., 1999, in `The Stellar content of the Local
Group', IAU Symp. 192, ed. P. Whitelock and R. Cannon\\
Hempel M., Kissler-Patig M., 2004, A\&A, 428, 459\\
Jarrett T.H., Chester T., Cutri, R., Schneider S., Skrutskie M.,
Huchra, J. P., 2000, AJ, 119, 2498\\
Jarrett T.H., Chester T., Cutri, R., Schneider S., Skrutskie M.,
Huchra, J. P., 2003, AJ, 125, 525\\
Kambas A., Davies J., Smith R., Bianchi S. and Haynes J., 2000. AJ, 120, 1316\\
Kauffmann G., White S., Guiderdoni B., 1993, MNRAS, 264, 201\\
Kauffmann G., et al., 2003, MNRAS, 341, 54\\
Klypin A., Kravtsov A., Valenzuela O., Prada E., 1999, ApJ, 522, 82\\
Kobayashi C., 2005, MNRAS, 361, 1216\\
Kodama T., et al., 2004, MNRAS, 350, 1005\\
Kuntschner H., 2000, MNRAS, 315, 184\\
Marigo P., Bressan A., Chiosi C., 1996, A\&A, 313, 545\\
Mateo M., 1998, ARAA, 36, 435\\
Mathis H., Lemson G., Springel V., Kauffman G., White S. D. M., Eldar A.,
Dekel A., 2002, MNRAS, 333, 739\\
Mannucci F., Della Valle M., Panagia N., Cappellaro E., Cresci G., Maiolino
R., Petrosian A., Turatto M., 2005, A\&A, 433, 807\\
Mart\'inez-Delgado D., Aparicio A., Gallart C., 1999, AJ, 118, 2229\\ 
Mendoza E. E., Johnson H. L., 1965, ApJ, 141, 161\\
Michard R., 2005, A\&A, 429, 819\\
Mighell K., Rich M., 1996, AJ, 111, 777\\
Moore B., Lake G., Quinn T., Stadel J., 1999, MNRAS, 304, 465\\
Norberg et al., 2002, MNRAS, 336, 907\\
Peletier R. F., Valentijn E. A., Jameson R. F., 1990, A\&A, 233, 62\\
Peletier R. F., Balcells M., 1996, AJ, 111, 2238\\
Persson S. E., Aaronson M., Cohen J. G., Frogel J. A., Matthews K.,
1983, ApJ, 266, 105\\
Phillipps S., Parker Q., Schwartzenberg J., Jones, J.,  1998a, ApJ, 493, L59\\
Phillipps S., Driver S. P., Crouch W. J., Smith R. M., 1998b, ApJ,
498, 119\\
Pietrinferni A., Cassisi S., Salaris M., Castelli F., 2004, ApJ, 612,
168\\
Poulain P., 1988, A\&AS, 72, 215\\
Poulain P., Nieto J.-L., 1994, A\&AS, 103, 573\\
Press W. H., Flannery B. P., Teukolsky S. A., Vetterling W. T., 1992,
``Numerical Recipes in C'', Cambridge University Press\\
Pritchet C. J., van den Bergh S., 1999, AJ, 118, 883\\
Puzia T. H., Zepf S. E., Kissler-Patig M., Hilker M., Minniti D.,
Goudfrooij P., 2002, A\&A, 391, 453\\
Reimers D., 1975, Mem. Soc. R. Sci.
Li\`ege, 8, 369\\
Rich R. M., Corsi C. E., Cacciari C., Federici L., Fusi Pecci F.,
Djorgovski S. G., 2005, AJ, 129, 2670\\
Roberts S., Davies J., Sabatini S., van Driel W., O'Neil K., Baes M., 
Linder S., Smith R., Evans R., 2004, MNRAS, 352, 478\\
Roberts et al., 2005, MNRAS, submitted\\
Sabatini S., Davies J., Scaramella R., Smith R., Baes M., Linder S. M., 
Roberts S., Testa V., 2003, MNRAS, 341, 981\\ 
Sabatini S., Davies J., van Driel W., Baes M., Roberts S., Smith R., 
Linder S., O'Neil K., 2005, MNRAS, 357, 819\\
Salpeter, E. E., 1955, ApJ, 121, 161
Sandage A., Wildey R., 1967, ApJ, 150, 469\\
Schlegel D. J., Finkbeiner D. P., Davis M., 1998, ApJ, 500, 525\\
Skillman E. D., Tolstoy E., Cole A. A., Dolphin A. E., Saha A.,
Gallagher J. S., Dohm-Palmer R. C., Mateo M., 2003, ApJ, 596, 253\\
Smail et al., 2001, MNRAS, 323, 839\\
Smecker-Hane T. A., Stetson P. B., Hesser J. E., Vandenberg D. A., 1996, 
in `Stars to galaxies: the impact of stellar physics on galaxy evolution', 
ASP Conf. Ser., Vol. 98, p. 328\\
Spergel D. N. et al., 2003, ApJS, 148, 175\\
Tantalo R., Chiosi C., 2004, MNRAS, 353, 405\\
Tully R. B., Somerville R. S., Trentham N., Verheijen M. A. W., 2002,
ApJ, 569, 573\\
van den Bergh S., 1981, A\&AS, 46, 79\\
Vassiliadis E., Wood P. R., 1993, ApJ, 413, 641\\
Vazdekis A., Arimoto N., 1999, ApJ, 525, 144\\
Wagenhuber J., Groenewegen M. A. T., 1998, A\&A, 340, 183\\
Wagenhuber J., 1996, PhD thesis, Techn. Univ. M\"unchen\\
Worthey G., Faber S. M., Gonzalez J. J., 1992, ApJ, 398, 69\\
Worthey G., 1994, ApJS, 95, 107\\
Wyder T.K., 2001, AJ, 122, 2490\\


\label{lastpage}

\end{document}